\documentclass[%
aip,
sd,%
amsmath,amssymb,
preprint,%
]{revtex4-1}

\usepackage{graphicx}
\usepackage{dcolumn}
\usepackage{bm}
\usepackage{siunitx}

\begin{document}
	
	\preprint{AIP/123-QED}
	
	\title{Jitter-calibrated second-order correlation measurement of low-repetition-rate pulsed excitation} 
	
	
	
	\author{Jue Gong}
	\affiliation{State Key Laboratory of Modern Optical Instrumentation, College of Optical Science and Engineering, Zhejiang University, Hangzhou 310027, China}
	
	\author{Hao Wu}
	\affiliation{State Key Laboratory of Modern Optical Instrumentation, College of Optical Science and Engineering, Zhejiang University, Hangzhou 310027, China}
	
	\author{Xin Guo}
	\affiliation{State Key Laboratory of Modern Optical Instrumentation, College of Optical Science and Engineering, Zhejiang University, Hangzhou 310027, China}

	\author{Wei Fang}
	\email[Email address: ]{wfang08@zju.edu.cn}
	\affiliation{State Key Laboratory of Modern Optical Instrumentation, College of Optical Science and Engineering, Zhejiang University, Hangzhou 310027, China}
	
	\author{Limin Tong}
	\affiliation{State Key Laboratory of Modern Optical Instrumentation, College of Optical Science and Engineering, Zhejiang University, Hangzhou 310027, China}
	
	
	\date{\today}
	
	\begin{abstract}
		We report a new technique for the realization of second-order correlation ($g^2(\tau)$) measurement under low-repetition-rate pulsed excitation (1 kHz), with timing jitter calibrated to restore lateral $g^2(\tau)$ curves and determine $g^2(0)$. We use CdSe nanowire (NW) laser to demonstrate the jitter-calibrated $g^2(\tau)$ measurement, where $g^2(0)$ evolution shows the laser emission transition process. The exciting pulses are split into reference and excitation channels, which enables the jitter calibration when the exciting pulses have significant timing jitter and low repetition rate. After using the reference signals to calibrate and rearrange the single-photon signals, lateral $g^2(\tau)$ curves can be entirely restored, and $g^2(0)$ evolution is demonstrated from our method.
	\end{abstract}
	
	
	\maketitle 

	\section{Introduction}
	
	Second-order correlation measurement performs excellent potential as an atomic-scale dynamics probe in light-matter interactions at the nanoscale\cite{1}. This measurement provides unique access to characterizing the dynamics of some critical physical processes, such as the charge transport\cite{2}, molecular motion\cite{3}, and quantum properties of light fields\cite{4,5,6,7,8,9,10,11,12}. In terms of a nanolaser, measuring the photon correlation function of the nanolaser’s output with a Hanbury Brown-Twiss (HBT) interferometer is a more excellent technique to explore lasing dynamics compared to the conventional L-L curve method\cite{13,14,15}, since some nanolasers with high spontaneous emission factors ({$\beta$}) show obscure lasing thresholds\cite{16,17,18}. Up to date, most photon correlation measurements are conducted under continuous wave (CW) excitation or high-repetition-rate pulsed excitation (over 1MHz)\cite{15,19}, in which the excessive energy of the shorter-wavelength pumping photons with respect to the output photons is converted to heat, especially in the relatively low-quantum-efficiency lasing system with high defect population and low quality of the resonant cavities\cite{20}. To avoid the thermal effect caused by heat accumulation or slow down the photobleaching process, researchers prefer low-repetition-rate pulsed excitation with short pulse duration\cite{21,22,23,24,25,26}, in which the photon correlation measurements need to work well with low-repetition-rate pulsed light fields. However, most low-repetition-rate excitation systems experience serious jitter effects (e.g., passively Q-switching laser), especially when the pulse duration is less than the timing jitter, leading to a severe distortion on the lateral $g^2(\tau)$ curves\cite{27}.
	
	Here we propose a novel idea to drastically reduce the jitter effects in photon correlation measurements by calibrating and rearranging the single-photon signals with a reference channel split from excitation pulses. Since an extra photon detector can record all the time intervals and the locations of the excitation pulses, single-photon signals will be calibrated and rearranged to eliminate the noise and timing jitter. Furthermore, low-repetition-rate excitation systems usually have low duty cycles (less than $1\%$), in which the time intervals between pulses are much larger than the pulse widths. Thus, we must find a lateral $g^2(\tau)$ pulse curve on a large time scale to normalize $g^2(0)$\cite{9}.
	
	Owing to their excellent properties to work as waveguide lasers with miniaturized footprints\cite{20,28}, we use CdSe nanowire (NW) laser as light source to prove that the lateral $g^2(\tau)$ curves can be restored, and jitter-calibrated photon correlation measurements will be achieved through our method.

	\section{Experimental Set-up and Method}
	
	Second-order correlation measurement towards nanolasers is growing popular in coherence characterization and estimation of the lasing threshold\cite{15}. Different photon statistics and coherent time determine the $g^2(0)$ and the $g^2(\tau)$ evolution for a continuous light wave. Theoretically, in a laser with the increasing intensity of pump power, the light field emitted from the nanolaser transitted from chaos to coherent field leads to decreasing $g^2(0)$ from 2 to 1. Whereas, experimentally, we can only find that the $g^2(0)$ increases from 1 to a particular value between 1 and 2 firstly, and finally decreases to 1, due to the limited time resolution of the single-photon detector\cite{29}. In particular, for a pulsed light wave, the $g^2(\tau)$ curve shows pulsed shapes with the same repetition rate of the pulsed light. Thus, the timing jitter of pump pulses will impact $g^2(\tau)$ calculation, leading to a severe distortion on the lateral $g^2(\tau)$ curve when the pulse jitter is larger than the pulse width\cite{27}.
	
	\begin{figure}[h]
		\includegraphics[width=6in]{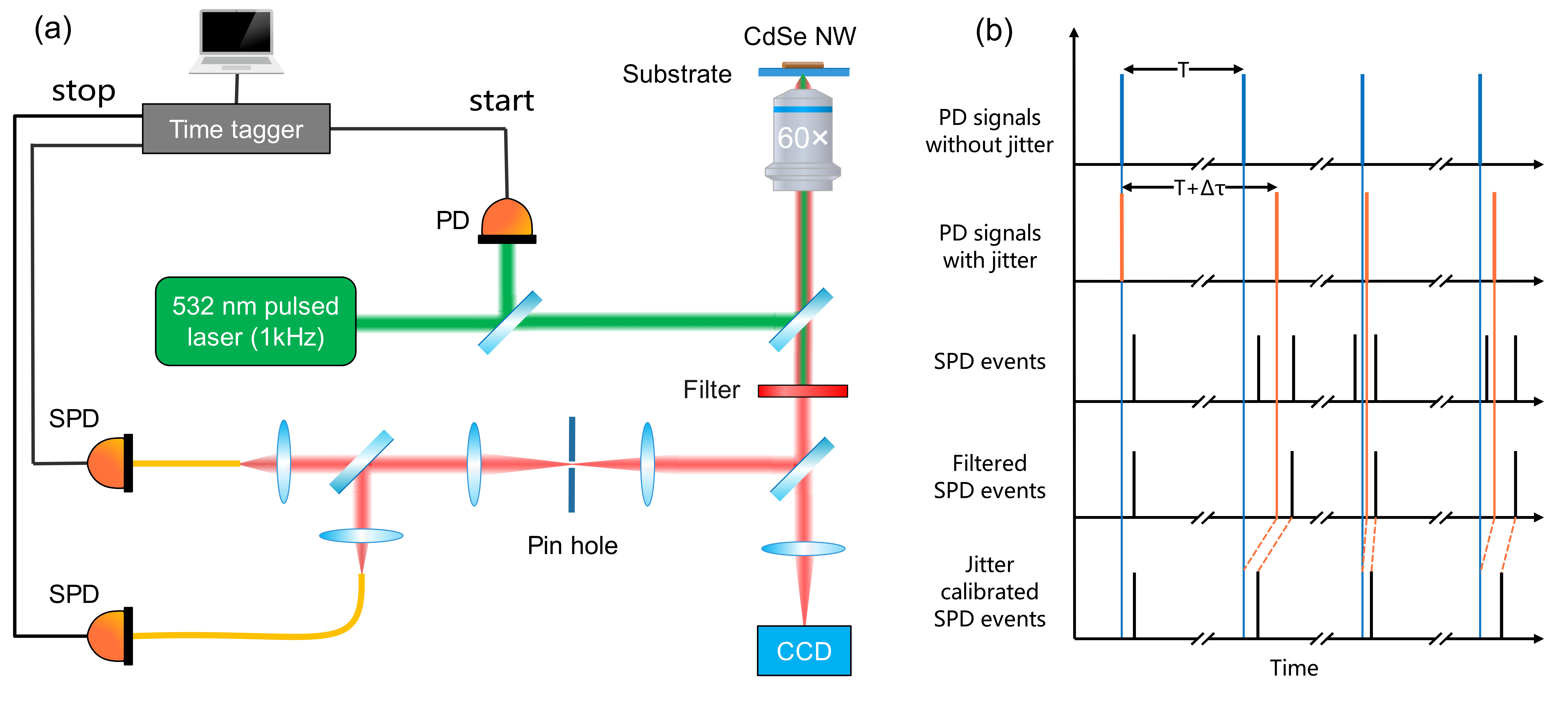}
		\caption{Jitter-calibrated second-order correlation measurement of a pulse-excited single CdSe NW. (a) Experimental set-up diagram of the jitter-calibrated second-order correlation measurement system. (b) Schematic diagram of the jitter-calibrated method. The blue lines show PD signals without jitter, corresponding to the start position of jitter calibrated SPD events. The orange lines show PD signals with jitter as time probe to calibrate all single-photon events and record all time intervals. Finally, we can rearrange all SPD events and calculate the jitter-calibrated $g^2(\tau)$.}
	\end{figure}
	
	The experimental set-up for our photon correlation measurement is illustrated in Fig. 1(a), in which the excitation system has a low repetition rate of 1 kHz, and the pulses are split into two channels before exciting CdSe NW. The optical emission is fed into two fibers using a beam splitter and detected by a pair of single-photon detectors (SPD) operated in the photon counting mode. The regime of signal processing is shown in Fig. 1(b), where the timing jitter of exciting pulses is calibrated through the reference photon detector (PD). For an ideal pulse exciting system, the pulse interval is considered to be a constant. However, timing jitter caused by exciting sources (e. g. chopper jitter or parasitic capacitance induced jitter) will disrupt the distribution of pulses, leading to a poor coincidence counting result. Thus, we use the reference channel to record all jitter information and make it a time start to calibrate all SPD signals out of the jitter. Finally, we rearrange all the SPD signals back to their expected position by subtracting their corresponding timing jitter, drastically improving the coincidence when calculating the lateral $g^2(\tau)$ curves.
	
	\begin{figure}[h]
		\includegraphics[width=3in]{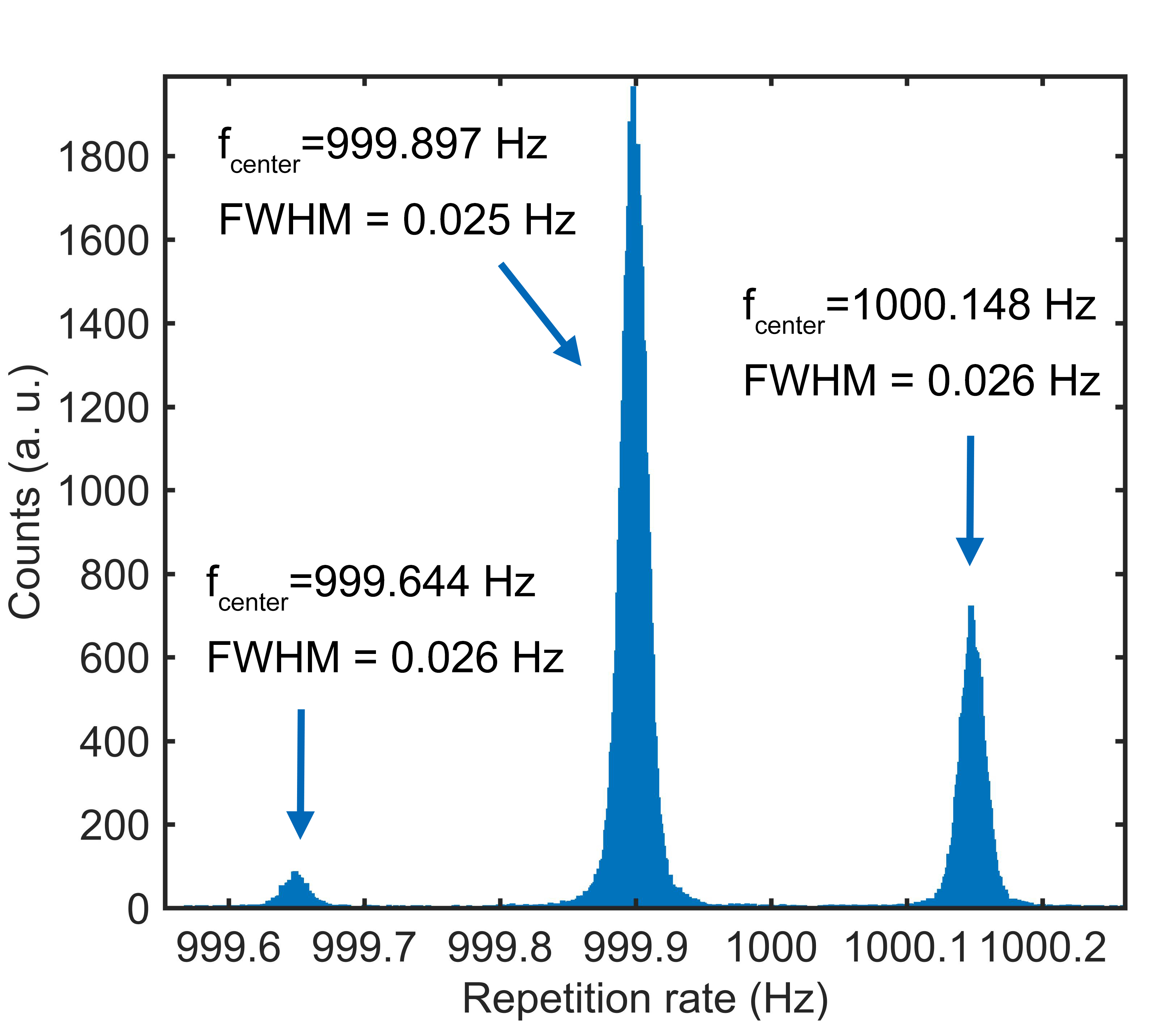}
		\caption{Repetition distribution of exciting pulses, where the central peak is located at 999.897 Hz, with corresponding FWHM 0.025 Hz. Two extra sidebands can be found due to the clock jitter of the electrical circuits in the exciting light source.}
	\end{figure} 
	
	Figure 2(a) shows the time distribution properties of exciting pulses, where we intend to use 1-kHz-repetition-rate pulses to excite the CdSe NW in 1 minute. However, the repetition rate of exciting pulses is not precisely located at 1 kHz. Two extra sidebands ($f_{center}=999.644\;\mathrm{Hz}, 1000.148\;\mathrm{Hz}$) can be found near the central peak of repetition rate distribution since the electro-optical modulator of the exciting source is strongly affected by the clock jitter of electrical circuits.
	
	\begin{figure}[h]
		\includegraphics[width=6in]{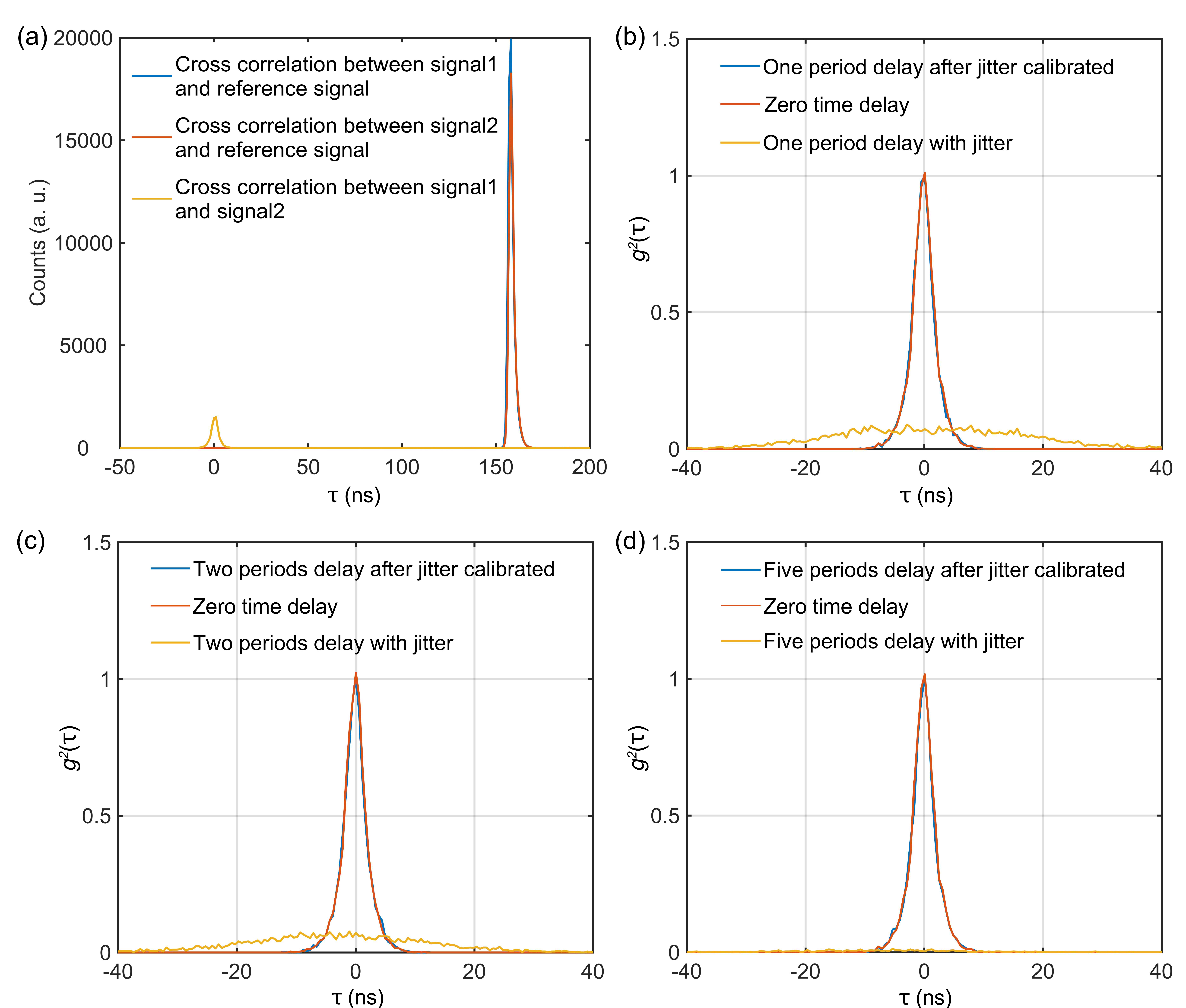}
		\caption{(a) Synchronization between SPD channels and PD reference channel. $g^2(\tau)$ curves are shown with (b) one period delay, (c) two periods delay, (d) five periods delay. The red line in (b)-(d) represents $g^2(\tau)$ under zero time delay (central $g^2(\tau)$). The blue and orange lines in (b)-(d) represent $g^2(\tau)$ under different periods delays with and without jitter calibrated, respectively. All the red and orange lines in (b)-(d) are translated to the zero position and normalized with the peak value of the corresponding blue line.}
	\end{figure}
	
	In order to calculate the second-order correlation of the pulsed light field, the common zero time reference needs to be found for the first step. Here we calculate the cross-correlation between reference and signal1, reference and signal2, signal1 and signal2, respectively. Since we can only get the waveform output from the PD/SPD channel, we use the Time Tagger 20 (Swabian instruments) to record the rising and falling edges of all the waveforms and choose the central time as the time-tag signal. Reference signal represents the time-tag signal from the reference PD channel, Signal1 and signal2 represent the time-tag signal from two SPD channels. In Fig. 3(a), both signal1 and signal2 show cross-correlation peaks near 160 ns with the reference signal, which means that two SPDs have around 160 ns time delay from reference PD.
	To investigate the jitter-calibrated $g^2(\tau)$ curve evolution with different time delays, we excite the CdSe NW to lasing state (corresponding to Fig. 5(e)), where the $g^2(0)$ is expected to be 1. Figure 3(b)-(d) shows $g^2(\tau)$ curves with 1, 2, 5 pulse periods time delays. When jitter calibrated, the lateral $g^2(\tau)$ curves (several periods time delay) fit the central $g^2(\tau)$ (No time delay) well and the $g^2(0)$ maintains 1, which is consistent with the photon correlation result under lasing condition. However, without jitter calibrated, the lateral $g^2(\tau)$ curve (orange curve) becomes weaker and weaker compared to central $g^2(\tau)$, and $g^2(0)$ deviates from 1 further and further away.
	 
	\section{Result and Discussion}
	
	\begin{figure}[h]
		\includegraphics[width=6in]{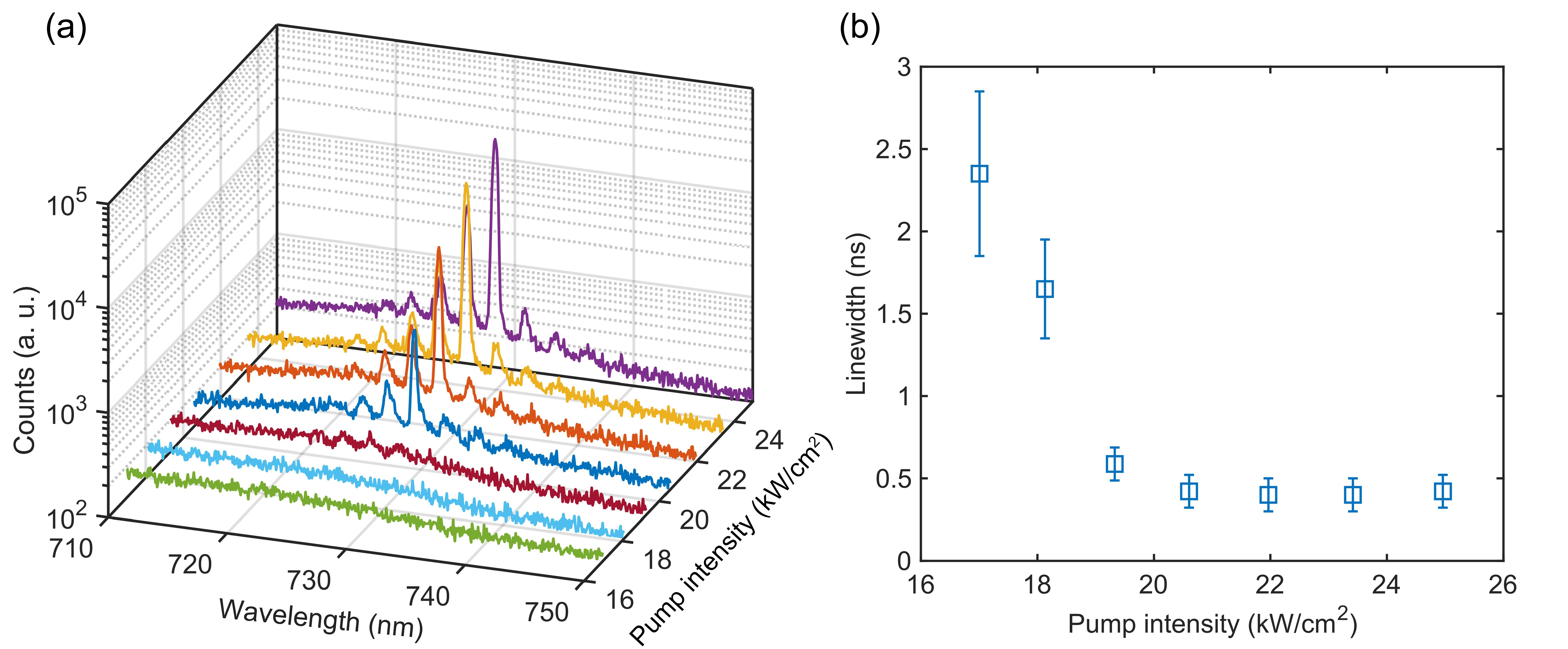}
		\caption{(a) Spectral evolution of the single CdSe NW laser. (b) The pump-dependent linewidth of the single CdSe NW laser.}
	\end{figure}
	
	\begin{figure}[h]
		\includegraphics[width=3in]{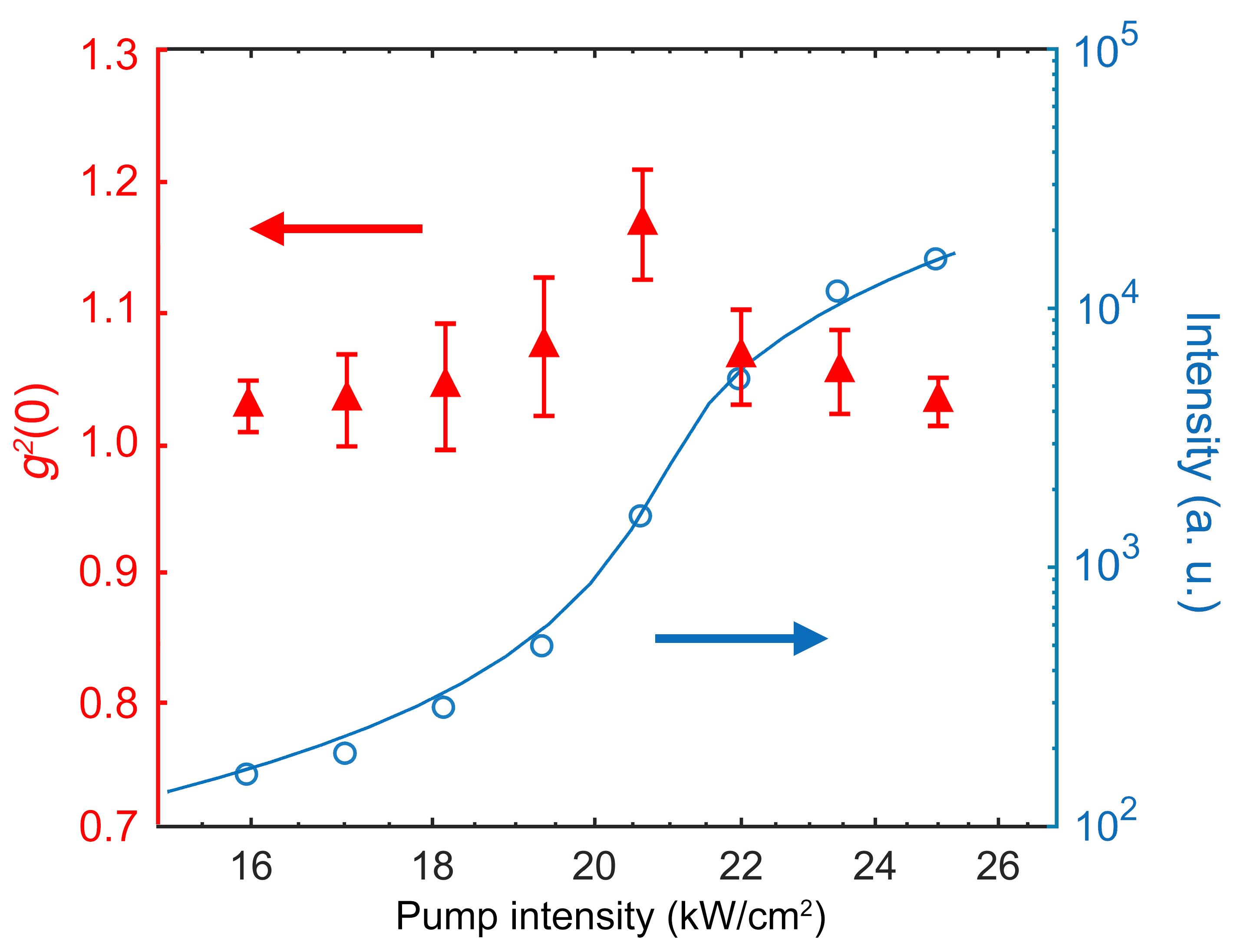}
		\caption{Pump-dependent $g^2(\tau)$ evolution of the single CdSe NW laser. The pump-dependent spectral intensity is shown with the blue circles and fit with the blue line. The red triangles represent the average $g^2(0)$, where we choose -5 to 5 periods time delay to calculate the mean values and standard deviations.}
	\end{figure}

	The CdSe NWs used in this work, with typical diameters of ~300 nm and length of 20 $\mu$m, were synthesized by the physical vapor deposition method\cite{30}. To investigate the lasing transition process, a CdSe NW was transferred onto the surface of a coverslip, a 532 nm pulsed laser with 1 kHz repetition rate was used as the excitation source, and the pump intensity of the nanowire was controlled using an adjustable attenuator, respectively. Figure 4(a) shows the emission spectrum of the NW lasing transition with the dominant mode peak around 728.4 nm, and Fig. 4(b) gives the linewidth evolution of the central peak in the emission spectrum, where the linewidth decreases to 0.45 nm when the NW is pumped above the lasing threshold.
	
	To show second-order correlation measurement with low-repetition-rate excitation, we calculate $g^2(\tau)$ curves of the CdSe NW at different pump intensities in Fig. 5. We use two photon counting modules (EXCELITAS, SPCM-AQRH-15-FC) as SPDs with dark counting rate $<$ 50 Hz, and Time Tagger 20 as signal collection board. Due to the low photon-counting rate ($\sim100$ counts/s) and low average photon number($\sim0.05$ counts/pulse), the measurement time is set to 10 minutes to collect enough SPD data, and the SPD signals are far less than the reference signals, as shown in Fig. 3(a).
	
	The $g^2(0)$ of the CdSe NW emission light field as a function of the pump intensity is presented in Fig. 5. With the increasing pump intensity on NW, $g^2(0)$ rises from 1 to 1.2 since $g^2(0)$ can only be resolved if the coherence time of the light field exceeds the detector resolution (350 ps), where the transition from thermal emission to lasing is accompanied by an increase in coherence time. After pump intensity reaches maximum point, $g^2(0)$ falls back to 1 since the light emission from NW gradually evolves into the laser. However, without the jitter-calibrated process, the lateral $g^2(\tau)$ curves will behave as the orange lines shown in Fig. 3(b)-(d), where a severe distortion occurs, and we can no longer make normalization to find the $g^2(0)$.
	
	\section{Conclusion}
	
	In conclusion, we have demonstrated a novel approach to calibrate the jitter effects in second-order correlation measurement, which is achieved via a reference channel locating, calibrating, and rearranging the SPD signals at a low-repetition-rate excitation. Compared to the previous experiments usually conducted at high-repetition-rate excitations and neglected the jitter effects, the jitter-calibrated method may greatly facilitate the low-repetition-rate exciting systems and relieve the heat accumulation in devices with low thermal damage thresholds. Meanwhile, many dye nanolasers or organic gain materials have been realized in recent years, and our method provides a unique way to correlation measurements while slowing down the photobleaching process or material lifetime. Also, the lateral $g^2(\tau)$ curves can be entirely reconstructed to normalize $g^2(0)$ so that the jitter of excitation sources will no longer impede our research on the laser transition process. More generally, the jitter-calibrated second-order correlation measurement allows low-repetition-rate excitation that is impossible to calculate $g^2(0)$ under serious jitter effects, paving the way toward low-repetition-rate excitation laser dynamics and quantum laser applications.
	
	%
	
	%
	
	\begin{acknowledgments}
		We thank Dr. Yixiao Gao, Mr. Peizhen Xu and Miss Luqing Shao for their great help in guidance and discussions. This work was supported by the (FUNDATIONS...)
	\end{acknowledgments}
	
	\bibliography{Reference}
	
\end{document}